# Electrodynamics of Nearly Ferroelectric Superconductors in the non-local Pippard limit.

U. Aparajita[1,2], J. Birman[1], O. Roslyak[3]

[1]*The City College, CUNY , 160-Convent Ave., New York, NY,10037.*
[2]*Queensborough Community Colleg, CUNY, 222-05 56th Ave., Bayside, NY, 11364.*
[3]*Hunter College, CUNY, 695-Park Ave., New York, NY, 10065.*

We report the structure of the magnetic field and secular current in a Nearly Ferroelectric Superconducting (NFE-SC) thin film. It was shown that unlike in conventional superconducting films, the external radiation causes alternating pattern of current strips. The strength of the innermost current torrents is governed by the laser field intensity as well as resonance with the ferroelectric component. The latter is modeled by secular reflection and random scattering in the Pippard non-local limit. Our calculations suggest that corresponding magnetic field pattern affects vortex formation in such material.